\pdfoutput=1
\documentclass{article}

\usepackage{arxiv}
\usepackage{graphicx}
\graphicspath{{figures/}}

\usepackage{booktabs}

\title{Student Use of LLMs and the Limits of AI-Generated Question Difficulty in Data
Science Courses}

\author{
 Yuan An \\
  School of Computer and Information Sciences\\
  Nick Howley College of Engineering and Computing\\
  Drexel University\\
  Philadelphia, PA 19104 \\
  \texttt{ya45@drexel.edu} \\
     \And
 Lei Wang \\
  School of Computer and Information Sciences\\
  Nick Howley College of Engineering and Computing\\
  Drexel University\\
  Philadelphia, PA 19104 \\
  \texttt{lw474@drexel.edu} \\
 }
 
 \date{}
 
\begin{document}
\maketitle

\begin{abstract}
This paper presents a multi-source classroom study conducted during 
a 10-week quarter in data science courses at Drexel University. 
We first investigate the behaviors of student engagement 
with large language models (LLMs) using four surveys across three data science courses. 
Second, we evaluate the construct validity of multiple-choice questions (MCQs) 
generated by an LLM for in-lecture retrieval practice. Based on 378 
authored questions (311 deployed, producing 7{,}888 student responses), 
we analyze whether the difficulty ratings assigned by an LLM match empirical item difficulty.
Our study shows that student engagement with LLMs varied across courses and increased over the 
term. Although students expressed high satisfaction and reported saving 
considerable time, their perception of deep learning benefits declined, 
and many noted a tendency toward over-reliance. Regarding the difficulty ratings
of LLM-generated MCQs, the Easy, Medium, and Hard labels correlated 
closely with its assigned Bloom's Taxonomy levels (Spearman $\rho=0.90$), 
reflecting an artifact of co-generation. However, neither metric predicted 
empirical item difficulty (difficulty label $\rho=0.06$; Bloom level $\rho=0.02$). 
The ratings reflect the structural formatting of a question rather than its underlying difficulty. 
\end{abstract}

\section{Introduction}
\label{sec:intro}

\begin{figure}[t]
  \centering
  \includegraphics[width=\columnwidth]{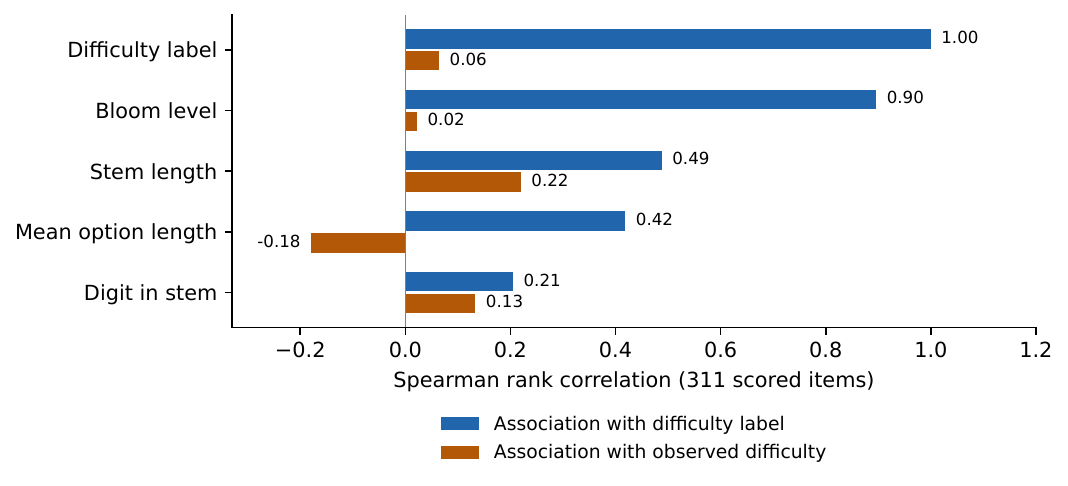}
  \caption{For each feature of an LLM-generated question we plot its Spearman
    correlation with the LLM's self-assigned difficulty label (blue) and with
    \emph{empirical} item difficulty measured from student responses
    (orange). The first blue bar reaches 1.00 because the difficulty label is
    correlated with itself. That bar serves as a reference point.
    The difficulty label itself is uncorrelated
    with how hard students find items ($\rho=0.06$). Other form features like length 
    only weakly predict empirical difficulty. The option length even negatively predicts 
    the empirical difficulty ($\rho=-0.18$).}
  \label{fig:gap}
\end{figure}

Large language models (LLMs), including general assistants like ChatGPT, 
Claude, and Gemini, alongside coding tools such as Copilot and Cursor, 
are now routinely used in student coursework~\cite{Kasneci2023-chatgpt,wang2024-LLM-education}. 
Instructors encounter these tools in two distinct positions. They need to monitor 
and adapt to student usage patterns that can either aid or hinder learning. 
At the same time, LLMs offer instructors an efficient way to generate frequent 
retrieval practice, an established method for supporting long-term memory retention~\cite{roediger2011critical,agarwal2021retrieval}, without requiring 
manual question authoring. An LLM prompted by an instructor can produce retrieval-practice items, 
such as multiple-choice questions (MCQs), alongside metadata including difficulty 
ratings (Easy, Medium, Hard) and cognitive classifications based on Bloom's Taxonomy. 
Prior research indicates that LLM-generated MCQs can support classroom retrieval 
practice and improve learning outcomes~\cite{yusof2025chatgpt,an2025enhancing}. 
Whether the difficulty ratings generated by these models reliably reflect true 
item difficulty remains an open question~\cite{doughty2024comparative,biancini2024multiple}.

This paper presents a multi-source classroom study conducted over a 10-week academic 
quarter. We examine two main questions. First, we investigate how data science students 
use and perceive LLMs across different courses. Second, we evaluate whether LLM-assigned 
difficulty ratings correspond to empirical item difficulty. The findings regarding 
the second question, illustrated in Figure~\ref{fig:gap}, show a clear discrepancy. 
The difficulty ratings produced by the LLM correlated with the Bloom's Taxonomy level the 
model assigns alongside them. The ratings also correlate with question length and formatting. 
However, the Spearman correlation between LLM-assigned difficulty and empirical item 
difficulty was negligible ($\rho=0.06$). While the model applies a consistent 
internal framework when assigning difficulty, this framework reflects how questions 
are written rather than how challenging students find them.

This paper offers the following main contributions:
\begin{enumerate}
  \item A \textbf{multi-source observational study}. We track student LLM usage across 
  three data science courses (machine learning, big data and cloud computing, and 
  applied machine learning) over a 10-week quarter, using multi-wave surveys.
  \item A \textbf{construct-validity audit of LLM-generated MCQ difficulty}. 
  We evaluate 378 authored questions (311 deployed, yielding 7{,}888 student responses), 
  showing that LLM difficulty ratings depend systematically on the co-generated Bloom 
  level and on item formatting. The ratings fail to predict empirical item difficulty 
  ($\rho\approx0.06$). 
\end{enumerate}

The remainder of the paper is organized as follows. Section~\ref{sec:background} 
reviews related work on student LLM usage, retrieval practice, and AI-generated 
assessments. Section~\ref{sec:context} describes the academic setting. Section~\ref{sec:methods} 
details the study design and methodology. Section~\ref{sec:results} presents findings 
from both the student usage investigation and the validity audit. Section~\ref{sec:discussion} 
addresses the broader implications of these results. Section~\ref{sec:conclusion} 
concludes with directions for calibrated retrieval practice systems.

\section{Background and Related Work}
\label{sec:background}

\subsection{LLM Use in Computing Education}
Adoption of LLMs in computing programs is high and continues to grow. In one 
project-based course from 2024, 80.5\% of students reported using LLMs, with 
ChatGPT being the most common tool~\cite{keuning2024students}. Automated code 
generation represents the primary use case across most studies, which rely mainly on self-reported survey data~\cite{prather2024beyond,becker2023programming}. Reported activities include 
writing code, debugging, understanding concepts, and verifying results, though 
novice students perform verification least often~\cite{leinonen2023comparing}.

Student over-reliance remains a significant concern. Beginner programmers often 
display passive behavior when interacting with AI code suggestions~\cite{prather2023weird}. 
This behavior can create an unearned sense of confidence that widens performance differences between students~\cite{prather2024widening}. Discrepancies also appear between 
self-reported AI usage and actual behavior~\cite{hak2025observinng}. 
A underlying factor is that beginner students often 
struggle to evaluate whether AI output is correct~\cite{zastudil2023generative}. 
Evidence regarding learning outcomes highlights potential risks. Relying on immediate 
AI suggestions can yield short-term task success while reducing conceptual 
transfer~\cite{kazemitabaar2023studying}. Students who relied on LLMs retained 
less material over time than those using web search engines~\cite{kumar2025togoogle}, 
leading instructors to adjust their teaching methods~\cite{lau2023ban}. 
Prior technical experience, rather than familiarity with AI tools, serves as a 
stronger predictor of whether a student benefits from LLM 
assistance~\cite{ma2025noteveryone,garg2024analyzing,bo2025whosleader}. 
This variation motivates our multi-course study design described in Section~\ref{sec:methods}.

\subsection{Retrieval Practice and Formative Assessment}
Retrieval practice, also known as the testing effect, is a supported concept in 
learning science. Actively recalling information strengthens long-term retention 
more effectively than re-reading 
material~\cite{Roediger2006-test-enhanced,roediger2011critical,agarwal2021retrieval}. 
Low-stakes classroom practice through clickers or polls increases engagement and 
retention~\cite{Lyle2011-retrieving}. Recent research demonstrates that LLM-supported 
retrieval practice can improve exam performance~\cite{yusof2025chatgpt}. Generating 
high-quality questions at the frequency required for regular classroom practice 
demands significant manual effort, prompting interest in LLM-based generation.

\subsection{LLM-Generated Assessment Items and the Difficulty-Calibration Problem}
Researchers have used LLMs to generate multiple-choice questions, finding that 
intrinsic quality scores based on rubrics can measure human-authored questions~\cite{olney2023generating,doughty2024comparative,biancini2024multiple}, 
including alignment with Bloom's Taxonomy~\cite{elkins2024howteachers}. 
Most evaluations focus on how well a question is written rather than whether 
its metadata is accurate. Studies comparing generated questions directly against 
student response data suggest that students sometimes score lower on LLM-authored 
items despite positive expert reviews, pointing to potential miscalibration in item difficulty~\cite{witsken2025llms}.

One approach for improving generated question quality involves conditioning model 
outputs on a knowledge graph (KG) constructed from lecture content~\cite{an2025ratedistortion}. 
In prior work, KG-conditioned questions scored higher on rubric criteria than those 
generated from raw text alone~\cite{an2026scaling}. Because those evaluations relied 
on expert rubrics, empirical difficulty calibration remained unexamined. The present 
study pools all generated items without evaluating KG conditioning directly, though 
knowledge graph structures are revisited in Section~\ref{sec:conclusion} as a framework 
for empirical calibration.

The central issue examined here is construct validity. Whether an LLM's self-assigned 
difficulty ratings and cognitive levels predict student performance in an active 
classroom setting. We evaluate this using Classical Test Theory (CTT). Metrics such 
as the difficulty index ($p$-value) can be applied 
to small classroom datasets~\cite{lordnovick1968,crockeralgina1986}.

\section{Course Context}
\label{sec:context}

This study was conducted over a 10-week academic quarter at Drexel University 
across three data science courses taught by two instructors. Table~\ref{tab:courses} 
outlines the participating courses.

\begin{table}[!h]
  \centering
  \caption{The three data science courses in the study (one 10-week quarter). 
  Enrollment is approximate roster size; per-instrument
    participation is a subset reported in Sections~\ref{sec:methods}
    and~\ref{sec:results}. UG: undergraduate.}
  \label{tab:courses}
  \small
  \begin{tabular}{@{}llcl@{}}
    \toprule
    Course & Level & $\approx$\,N & Instruments \\
    \midrule
    Course-ML  & Middle-level UG   & 20  & MCQs, surveys \\
    Course-BD  & Upper-division UG & 40  & MCQs, surveys \\
    Course-AML & Graduate          & 35 & Surveys only \\
    \bottomrule
  \end{tabular}
\end{table}

\textbf{Course-ML} is an intermediate undergraduate course covering logistic 
regression, decision trees, ensemble methods, overfitting, regularization, and 
model evaluation. \textbf{Course-BD} is an advanced undergraduate course on big 
data and cloud computing, covering Spark architecture, Resilient Distributed 
Datasets (RDDs), lazy evaluation, directed acyclic graph (DAG) execution, 
shuffling, partitioning, and PySpark DataFrames. \textbf{Course-AML} is a 
graduate-level applied machine learning course.
In Course-ML and Course-BD, instructors included short retrieval practice sessions 
during every lecture throughout the 10-week term. Every 15 to 20 minutes, 
students completed timed Canvas quizzes containing LLM-generated questions.

\section{Study Design and Methods}
\label{sec:methods}

\subsection{Overview}
This observational study involved no experimental manipulation and received 
approval from the Drexel University Institutional Review Board. The research 
includes two tracks. \emph{Track A} examines student LLM usage patterns 
through surveys. \emph{Track B} evaluates the construct 
validity of difficulty ratings assigned by LLMs to generated retrieval-practice questions. 
Track A is descriptive, presenting summary metrics due to varying sample sizes across courses.

\subsection{Student LLM Usage Study}
\label{ssec:usage}

\subsubsection{Multi-wave surveys}
\label{sssec:surveys}
Data on student behavior were gathered using four anonymous Qualtrics surveys 
distributed across the quarter: an initial survey in Week 1, two mid-term surveys 
(Weeks 3 and 5-6), and a final survey in Weeks 9-10. The surveys covered Course-ML, 
Course-BD, and Course-AML, alongside responses from unidentified courses. Across all 
waves, 94 total responses were recorded (37 from Course-BD, 23 from Course-AML, 
16 from Course-ML, and 18 from other or unspecified courses). The initial survey 
assessed prior tool usage, frequency, primary purposes, confidence in assessing 
LLM outputs, perceptions of learning impact, and comfort levels. Subsequent surveys 
measured course usage frequency, primary tasks (coding, debugging, conceptual 
understanding, initial drafts, and solution checking), specific task frequencies, 
perceived learning effects, agreement ratings (time saved, over-reliance, depth of 
learning, and study habits), and open-ended feedback. Given the sample size, 
these survey data are analyzed descriptively.

\subsubsection{Analysis}
Survey responses were analyzed to summarize usage frequency and tasks across 
courses and survey waves. Cross-course comparisons were conducted using the 
final survey wave, which used a shared set of questions. Open-ended responses 
were categorized topically through manual review.

\subsection{LLM-Generated MCQ Retrieval Practice}
\label{ssec:mcq}

\subsubsection{Generation and deployment}
Multiple-choice questions were generated from lecture content in six-item sets: 
two Easy, two Medium, and two Hard questions. Generation alternated between 
\emph{GPT-5.4-mini} and \emph{Gemini-3.1-flash-lite}, with each model creating 
roughly half the items. Each question included an LLM-assigned difficulty label 
(Easy, Medium, or Hard). Questions were presented in sets of three or four per 
lecture, with two to three minutes allocated per set. Students do not see the 
difficulty ratings of the questions. Across the term, 
378 questions were authored (126 Easy, 126 Medium, and 126 Hard). Of these, 
311 were deployed in class, generating 7{,}888 individual student responses 
from 54 participants across 52 quizzes in Course-ML and Course-BD. All items 
were evaluated as a combined corpus without treating model source as an independent variable.

\subsubsection{Construct-validity analysis}
\label{sssec:cva}
Question difficulty was evaluated in two stages. \emph{Stage 1} examined the 
full authored set of 378 questions. Structural properties, text length, and 
cognitive classifications were extracted, including stem length, option length, 
option length variance, numerical content, and co-assigned Bloom's Taxonomy level. 
Spearman correlations were applied to discrete categories to identify factors 
separating the three difficulty levels. \emph{Stage 2} analyzed the 311 deployed 
questions. Feature values and LLM difficulty ratings were correlated with empirical 
difficulty ($1-p$, where $p$ represents the proportion of correct student responses) 
to test whether features associated with LLM labels predicted empirical difficulty.

\section{Results}
\label{sec:results}

\subsection{Student LLM Use}
\label{ssec:use-results}

\subsubsection{Baseline: prior experience}
\begin{figure}[tb]
  \centering
  \includegraphics[width=\columnwidth]{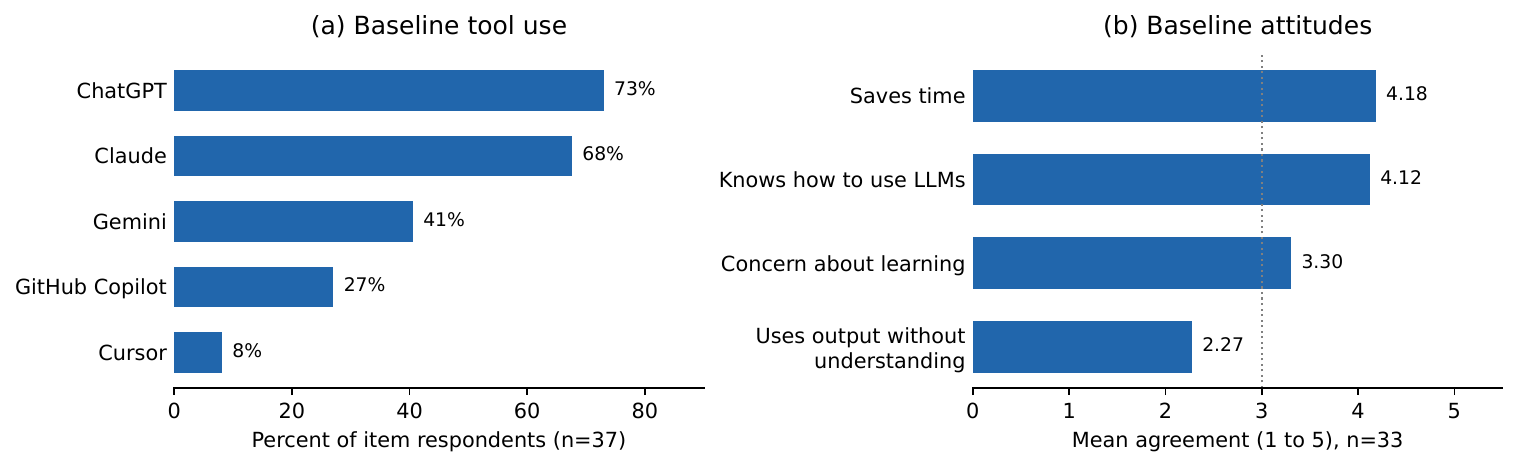}
  \caption{Baseline survey (Week~1). (a)~Prior tool prevalence among the 37
    students who answered the tool item. (b)~Baseline attitudes on a 1 to 5
    agreement scale, $n{=}33$ (dashed line marks the neutral midpoint).}
  \label{fig:baseline}
\end{figure}

Students reported high rates of LLM usage at the start of the course. ChatGPT and 
Claude were the most commonly reported tools in Week 1, named by 73\% and 68\% of the 
37 students who answered the tool item (Figure~\ref{fig:baseline}a). On a 1-to-5 
agreement scale, students reported that LLMs saved time and that they felt capable of 
using them effectively (Figure~\ref{fig:baseline}b). They reported lower rates of 
accepting AI outputs without review and expressed moderate concern regarding 
deep understanding. Across all 48 baseline respondents, prior usage was centered on 
debugging (75\%) and understanding concepts (67\%), followed by reviewing work (56\%) 
and starting assignments (33\%).

\subsubsection{In-course usage across waves}
\begin{figure}[tb]
  \centering
  \includegraphics[width=\columnwidth]{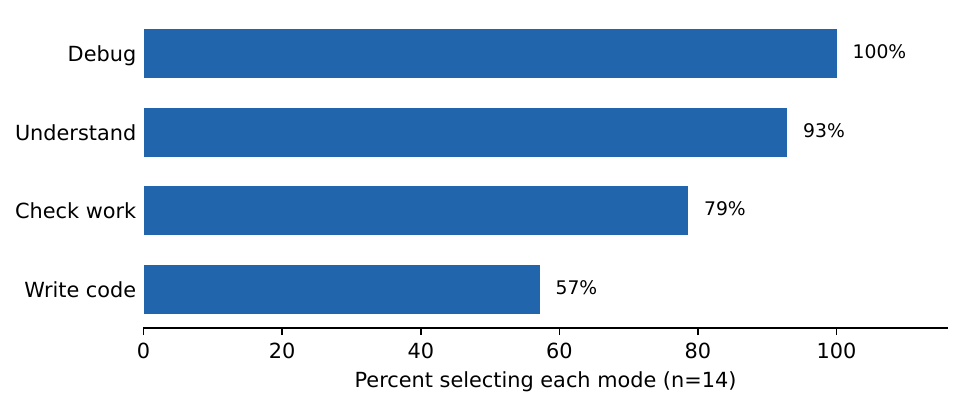}
  \caption{Debugging and concept questions were near universal.
  Code writing was the least common of the four shown.}
  \label{fig:modes}
\end{figure}

\begin{figure}[tb]
  \centering
  \includegraphics[width=\columnwidth]{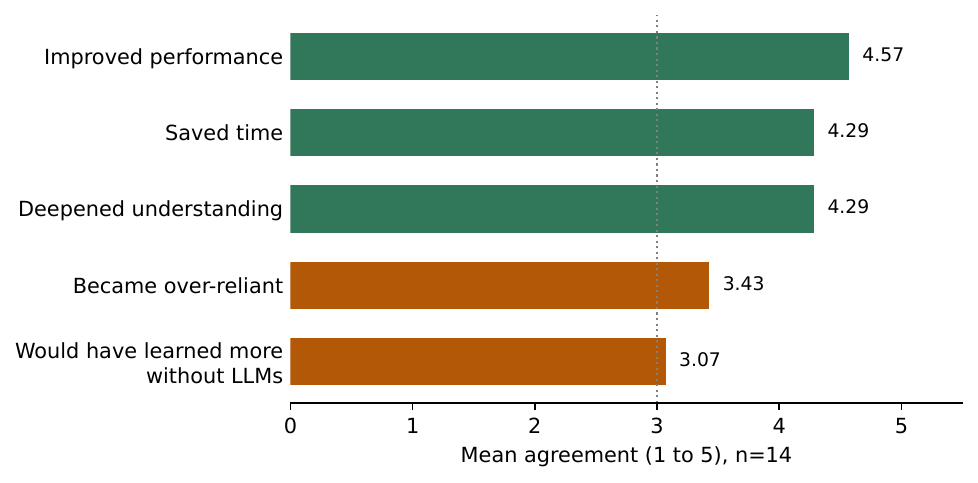}
  \caption{Mean agreement on the final survey (1 to 5, $n{=}14$). Students
    endorse the benefits (green) and report some over-reliance (orange). The
    last item sits at the neutral midpoint of 3. Students were divided about
    whether they would have learned more without LLMs.}
  \label{fig:agree}
\end{figure}

LLM adoption expanded over the quarter. By the final survey wave, using LLMs 
for debugging and understanding concepts was nearly universal 
(Figure~\ref{fig:modes}). 
Reported usage frequency reached its highest points in Course-BD and Course-AML 
(Figure~\ref{fig:course}a). Perceived benefits to learning declined over time. On 
a scale ranging from $-1$ (aiding only with task completion) to $+2$ (deepening 
conceptual understanding), the average score dropped from 0.67 in Week 3 to 0.47 
in the final survey, even as satisfaction remained high. The final agreement survey 
(Figure~\ref{fig:agree}) shows where students landed at the end of the term. Students 
noted improvements in performance, time savings, and comprehension. They also reported 
some over-reliance. The group was divided on whether they would have learned more 
without AI tools. That item averaged 3.07 on the 1-to-5 scale, at the neutral midpoint.

\subsubsection{Variation across courses}
\begin{figure}[tb]
  \centering
  \includegraphics[width=\columnwidth]{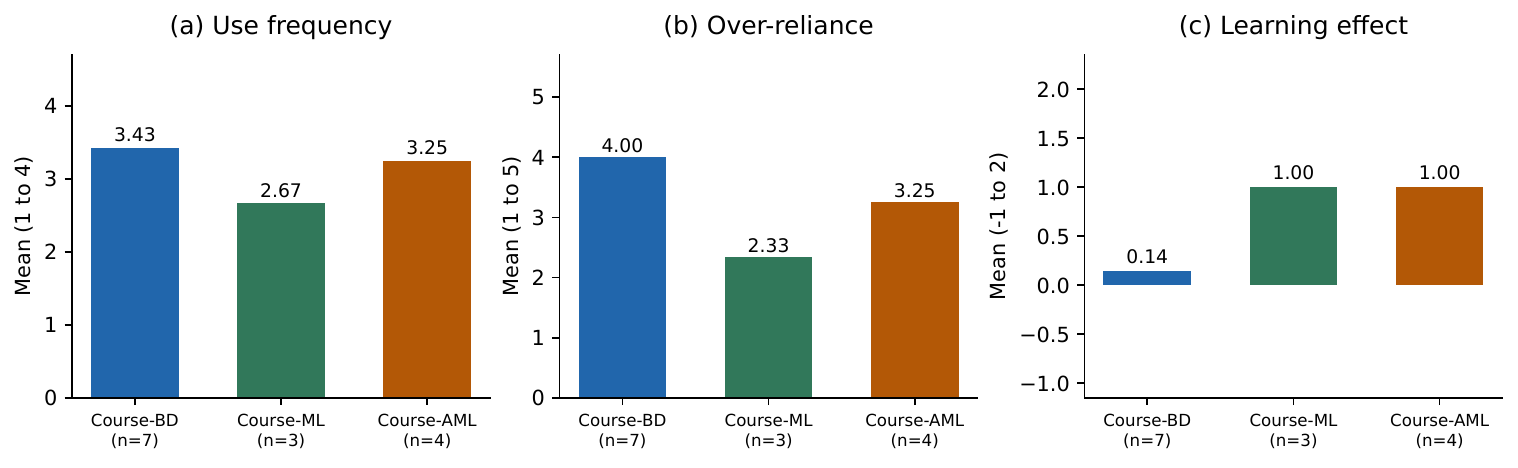}
  \caption{LLM use by course. (a)~Use frequency (1 to 4),
    (b)~self-reported over-reliance (1 to 5), and (c)~perceived learning effect
    ($-1$ to $2$). The scales differ across panels. Course-BD shows the highest
    frequency and over-reliance with a near-zero learning effect.}
  \label{fig:course}
\end{figure}

Course-BD showed the highest usage frequency, the highest self-reported over-reliance, 
and a perceived learning effect near zero (Figure~\ref{fig:course}). 
Course-ML exhibited lower usage frequency, lower over-reliance, and a positive 
perceived learning effect.
Course-AML showed usage frequencies similar to Course-BD alongside positive learning 
ratings similar to Course-ML.

\subsection{LLM-Generated Retrieval Practice: Deployment and Construct-Validity Audit}
\label{ssec:mcq-results}

\subsubsection{Deployment and performance}
Low-stakes retrieval practice items were delivered during lectures throughout the term.
Student scores were high overall, with approximately 70\% of items achieving a 100\%
correct response rate ($p=1.0$). This performance level restricts score variance, providing
context for the item statistics below.

\subsubsection{What the LLM treats as ``difficulty''}
\begin{figure}[tb]
  \centering
  \includegraphics[width=\columnwidth]{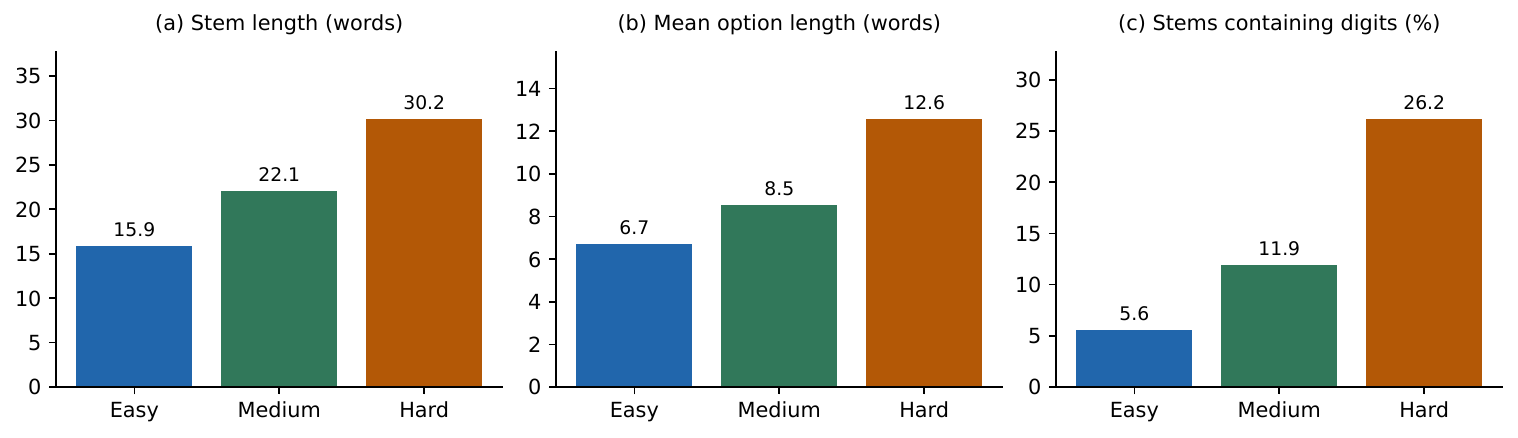}
  \caption{Item-form features that rise monotonically with the LLM's difficulty
    label across Easy, Medium, and Hard items. Bars cover the 378 authored
    items, 126 per label. (a)~Mean stem length in words. (b)~Mean option length
    in words. (c)~Percent of stems containing a digit.}
  \label{fig:feats}
\end{figure}

\begin{figure}[tb]
  \centering
  \includegraphics[width=\columnwidth]{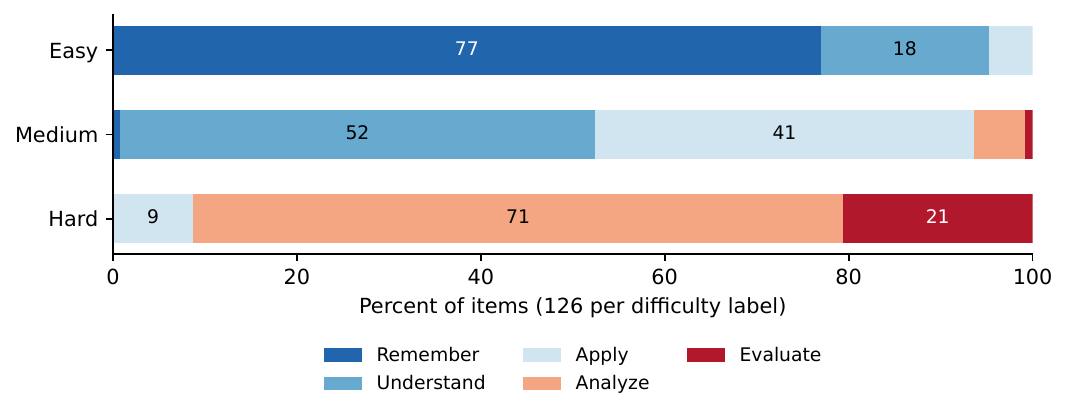}
  \caption{Bloom cognitive level composition of items at each LLM difficulty
    level. Bars stack to 100 percent, and small segments carry no label.
    Difficulty and Bloom are co-generated by the model, so this alignment
    reflects its labeling convention rather than item difficulty.}
  \label{fig:bloom}
\end{figure}

The difficulty labels assigned by the LLM correspond to structural question features. 
Question stem length, option length, and the presence of numerical digits increased 
systematically across Easy, Medium, and Hard categories (Figure~\ref{fig:feats}). 
Other surface characteristics, such as LaTeX formatting, code snippets, negation 
words, and numeric-option counts, did not separate difficulty levels. Assigned difficulty 
ratings correlated closely with co-generated Bloom's Taxonomy levels, moving from 
basic recall for Easy questions to analytical tasks for Hard questions 
(Figure~\ref{fig:bloom}). Over the 378 authored items, the difficulty rank and the 
Bloom rank gave a Spearman $\rho=0.90$ with $\chi^2$ $p\approx10^{-107}$. 
Because difficulty ratings and Bloom levels were generated together, this 
alignment reflects internal model prompting behavior rather than empirical difficulty. 
The assigned difficulty labels function primarily as a proxy for text length 
and model-selected cognitive tags.

\subsubsection{The construct-validity gap}
We evaluated whether these structural features predicted empirical question difficulty 
($1-p$), where $p$ is the proportion of correct student responses. As shown in 
Figure~\ref{fig:gap}, they did not. The LLM-assigned difficulty label showed no 
meaningful correlation with empirical difficulty ($\rho=0.06$), and assigned Bloom 
levels showed similar results ($\rho=0.02$). Structural features tracked by the model 
showed weak relationships with empirical difficulty. Option length ran in the opposite 
direction ($\rho=-0.18$). Items with longer answer choices were easier for students. 
The difficulty labels reflect formatting conventions rather than empirical difficulty.

\section{Discussion}
\label{sec:discussion}

\subsection{Student LLM Use in Data Science Courses}
Student usage trends revealed two clear patterns. First, usage patterns varied by 
course setting. Course-BD students engaged in more code generation and reported 
higher rates of over-reliance. Course-ML students reported lower usage overall 
and less code generation. Debugging was the top mode in both courses. While reported descriptively, this 
variation indicates that daily course requirements influence how students use 
AI tools. Second, students reported high satisfaction and time savings. They 
also noted a decline in deep conceptual learning. Reported over-reliance held 
steady across waves, a little above the midpoint of the scale. This outcome aligns with findings regarding 
learning retention in introductory programming environments~\cite{prather2024widening,kazemitabaar2023studying}.

\subsection{The Limits of AI-Generated Tests}
Integrating LLM-generated questions into regular lecture practice proved 
practical and was well received by students. However, model-assigned difficulty 
ratings did not reflect empirical item difficulty. Assigned difficulty levels 
reflect text length and co-generated cognitive tags rather than empirical 
item difficulty. Generated difficulty and Bloom's Taxonomy ratings should 
not be assumed to be calibrated. Instead, question difficulty can be evaluated 
using student response data. Realized difficulty depends on phrasing, specific terminology, and prior 
student knowledge rather than the target cognitive categories used by the 
model. 

\subsection{The Study Design and Limitations}
Combining survey data and item response metrics provided 
complementary insights into classroom implementation. However,  
this study has several limitations. Survey sample sizes were limited, data were 
gathered at a single institution, and classroom retrieval practice was managed 
by one instructor. Anonymous survey collection prevented tracking individual 
student changes over time, and high overall accuracy rates limited score 
variance. Accordingly, survey findings are presented as descriptive, and 
validity results are framed within the context of the analyzed question corpus.

\section{Conclusion and Future Work}
\label{sec:conclusion}

This study implemented LLM-generated retrieval practice 
in data science courses over a 10-week term, examining student usage 
patterns and evaluating model-assigned difficulty labels. Student AI usage 
varied across courses alongside reported trade-offs between efficiency and 
depth of understanding. Evaluated difficulty labels reflected structural 
item formatting rather than empirical item difficulty. These findings 
suggest that while LLMs can support question generation for lecture practice, 
assigned difficulty labels require empirical calibration based on student 
performance.

Future work could focus on developing adaptive retrieval practice systems that 
adjust question selection using student performance data. Estimating difficulty 
and discrimination values from student responses (using CTT initially and Item 
Response Theory as response volume grows) would allow systems to select, retire, 
or update questions to match target difficulty levels. Structuring lecture 
materials using a knowledge graph~\cite{an2025ratedistortion} provides a 
framework for generation, allowing systems to produce questions targeting 
specific concepts once target difficulty levels are defined. Combining 
knowledge graph generation~\cite{an2026scaling} with empirical calibration 
could help refine automated item generation. Additional areas for research 
include evaluating larger samples using Item Response Theory, examining 
implementation across multiple instructors, and conducting longitudinal 
studies on retention.

\end{document}